\documentclass[a4paper]{article}
\pdfoutput=1

\usepackage[utf8]{inputenc}
\usepackage[T1]{fontenc}
\usepackage[english]{babel}

\usepackage[pdftex,breaklinks=true,bookmarks=true,pdfborder={0 0 0},colorlinks=false]{hyperref}

\usepackage{listings}
\usepackage{amsfonts}
\usepackage{amsmath}
\usepackage{cleveref}
\usepackage{graphicx}
\usepackage[caption=false]{subfig}

\newcommand{\ie}{i.e.~}
\newcommand{\eg}{e.g.~}
\newcommand{\cf}{cf.~}
\newcommand{\vs}{vs.~}

\usepackage{enumitem}
\setlist{nolistsep}
\usepackage{array}

\begin{document}

\author{Raphael ‘kena’ Poss\\University of Amsterdam, The Netherlands}
\title{Extrinsically adaptable systems\\ \small or
  ``what are hacker-friendly systems and how we should ask for more''}

\maketitle

\begin{abstract}
  Are there qualitative and quantitative traits of system design that
  contribute to the ability of people to further innovate? We propose
  that \emph{extrinsic adaptability}, the ability given to secondary
  parties to change a system to match new requirements not envisioned
  by the primary provider, is such a trait. ``Extrinsic adaptation''
  encompasses the popular concepts of ``workaround'', ``fast prototype
  extension'' or ``hack'', and extrinsic adaptability is thus a
  measure of how friendly a system is to tinkering by curious
  minds. In this report, we give ``hackability'' or ``hacker-friendliness'' scientific
  credentials by formulating and studying a generalization of the
  concept. During this exercise, we find that system changes by
  secondary parties fall on a subjective gradient of acceptability,
  with extrinsic adaptations on one side which confidently preserve
  existing system features, and invasive modifications on the other
  side which are perceived to be disruptive to existing system
  features. Where a change is positioned on this gradient is dependent
  on how an external observer perceives component boundaries within
  the changed system. We also find that the existence of objective
  cost functions can alleviate but not fully eliminate this
  subjectiveness. The study also enables us to formulate an
  ethical imperative for system designers to promote extrinsic adaptability.
\end{abstract}

\clearpage

\setcounter{tocdepth}{1}
\tableofcontents

\clearpage

\section{Introduction}

Since the turn of the 21st century, powerful market forces are at play
to redefine the place of general-purpose computers. Before the last
decade, users could define the function of their computing devices
separately from form. They could do so by acquiring software, but also
by writing their own software, separately from the acquisition of the
platform. They could also replace hardware components by alternate
devices with the same interface but a different implementation.  In
contrast, the last decade has then seen the advent of tightly
integrated smart terminals and entertainment platforms, whose form and
function are bundled by the manufacturer and not separable by the
user. These devices have displaced commodity, modular general-purpose
platforms. They progressively erase the incentive to educate
non-technical audiences about the benefits of carrying out their own
innovation.

On the one hand, we could simply acknowledge this evolution as a
natural effect of a free market where consumers have decided, through
their purchasing power, their preference for pre-programmed fixed
functions and corporate control over features. On the other hand, we
could also worry about an opportunity loss, that of educating a larger
number of individuals to the power of modularity, reuse and
adaptability in computing systems. With the end of the ``free lunch''
ten years ago~\cite{ronen.01.ieee,sutter.05}, the industry is
struggling to devise new architectures and technical mindsets to
answer ever growing computational needs. It seems to us that a large
diversity of creative approaches by individuals empowered to innovate
will be needed to overcome these challenges.

Meanwhile, the words ``empowering to innovate'' may seem at first
sight to be a wilfull but empty shell. This author is dedicated to
give these words a concrete meaning in a scientific context. The
general line of research is to investigate the following question: are
there qualitative and quantitative traits of system design that
clearly contribute to or detract from the ability of people to further
innovate?

One specific trait that contributes to
innovation is the ability given to secondary parties to perform changes
to existing systems, to match requirements not initially envisioned by
the primary system provider. We call these changes \emph{extrinsic
  adaptations}. This is the process that a hobbyist uses when
replacing an engine carburetor by another to experiment with a
``greener'' alternative. It is the process by which many scientists
prototype their ideas of optimizations to existing systems. Extrinsic
adaptations, and the ability to perform them on existing systems,
constitute an essential \emph{instrument to innovation}. The rest of this
report thus focuses on how to describe
  extrinsic adaptations generally, and what properties of systems are
  favorable to extrinsic adaptations. These questions are important
because their answers provide guidelines to systems designers to
preserve and promote opportunities to innovate in the future.

We start in \cref{sec:def,sec:ex} by proposing a general definition
and a couple of illustrative examples. We then describe in
\cref{sec:extinv,sec:cost,sec:acc} the nature and impact of extrinsic
adaptations from the perspective of engineers and users. Strong with
this conceptual equipment, we then propose in \cref{sec:measure} some
general, quantitative measures of extrinsic adaptability, which we
then apply in the specific case of computing systems in
\cref{sec:compsys}. A summary is then provided in \cref{sec:conc}.

\section{The essence of extrinsic adaptation}\label{sec:def}

Any sufficiently complex system is an assembly of components where the
specification, provision, integration and exploitation of the
components are performed by and under the responsibility of different
parties.

When a new party comes in contact with an existing
system, they may have new expectations towards this system that were
not envisioned by the parties interacting with the system so far. From
the time these expectations are phrased into requirements, and until
the time where the entire system is changed to match these
requirements, a discussion occurs between the parties about whether
and how to best evolve the system. During this discussion, a process
of gradual change may occur where some components are adapted, and
where other components that should be modified are left unchanged for
non-technical reasons: money, intellectual property, lost source code,
politics, etc.

The study of these change processes is a branch of change management.  We
focus here on change scenarios where:
\begin{itemize}
\item a new requirement on a system is formulated by a secondary
  party, \ie not the primary parties providing the system;
\item components that should be modified by the primary parties to
  answer the new requirement are left unmodified;
\item instead, the system as a whole is coerced externally by a
  secondary party into adapting towards the new requirement, possibly
  in a way judged ``less than optimal'' by the primary parties.
\end{itemize}

We call this type of change \emph{extrinsic adaptation}. The term
encompasses other terms commonly used in more specific scenarios:
``workaround'', ``bypass'', ``stop-gap measure'', ``quick and dirty
solution'', ``temporary fix'', ``hack\footnotemark{}'', etc.

\footnotetext{using here the friendly meaning of the word ``hack'':
  ``a quick job that produces what is needed, but not well.'' \cf \url{http://www.hacker-dictionary.com/terms/hack}.}

\begin{figure}
\hspace{-.2\textwidth}\includegraphics[width=1.4\textwidth]{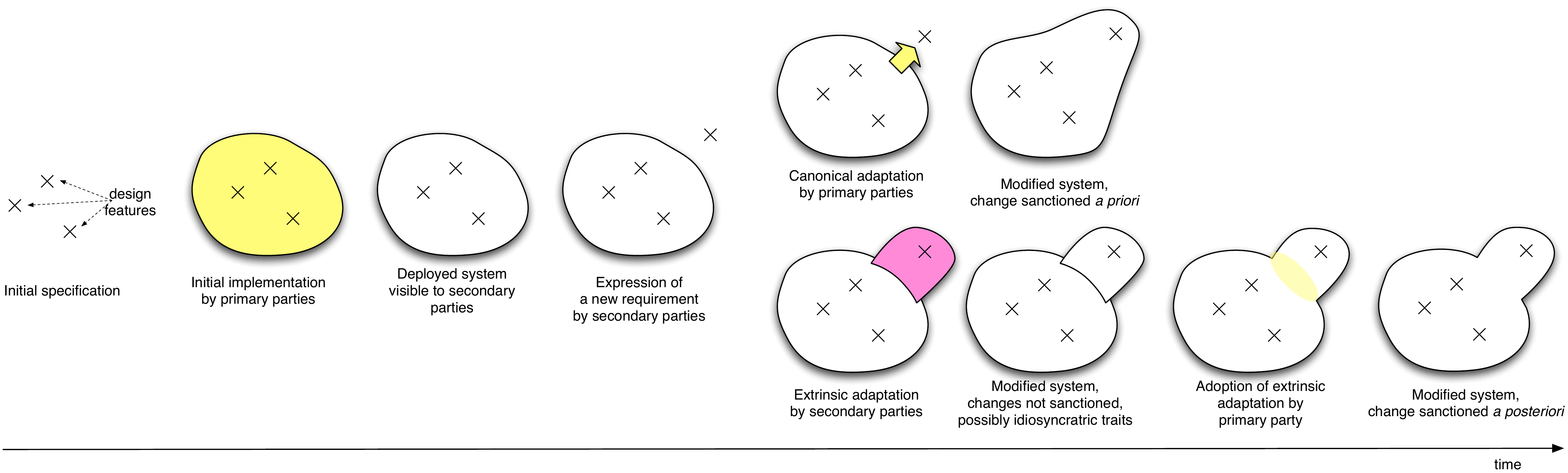}
\caption{Alternate change strategies to react to new requirements.}\label{fig:evolution}
\end{figure}

An extrinsic adaption exists from the moment a system is modified in
this way towards a new requirement. It stops to exist when the primary
parties either adopt the change, \ie extend their vision of the system
to include the adaptation as an intrinsic new feature, or adapt the
system in a different way towards the same requirement, \ie perform an
alternate, ``canonical'' adaptation. While the extrinsic adaptation
exists, it may (but does not have to) be perceived to be
idiosyncratic, somewhat ``out of line'' from the systems' overall
design guidelines and aesthetics. We illustrate these aspects in \cref{fig:evolution}.

In the rest of this report, we are primarily interested in component
boundaries defined by \emph{behavior interfaces} (``black box
model''), which specify which signals/actions can influence the
component's behavior and what are its observable effects on its
environment. For example, in the hardware domain, this
definition maps to physically bounded components, where behavior is
defined by the signals entering and leaving the component boundary; in
the software domain, it maps to programming interfaces (APIs) and what
side effects from the component are observable from the run-time
environment's perspective. Note however that we do not limit our argument to computing
systems until \cref{sec:compsys}.

\section{Example scenarios of extrinsic adaptation}\label{sec:ex}

The following sub-sections provide concrete examples of extrinsic
adaptation and highlight why they are desirable.

\subsection{Non-canonical user expectations in a video game}

Angry Birds from Rovio Entertainment is a video game running on
computers equipped with a touch-sensitive display. From the player's
perspective, the primary function of the game is to use birds as
projectiles, using finger gestures on the display, to dislodge green
pig faces hidden behind movable obstacles. At a high level, this
program is thus a function of the finger gestures as input, towards
graphical animations and the computation of score points as
output.

We consider the system formed by the device and the game program, both
being components interacting with each other.

Meanwhile, the free-as-in-beer version of this video game also
displays advertisements at the bottom or top of the display while the
game is ongoing. The display of advertisements is mostly
inconsequential to the players' experience, and usually ignored by
them, but owners of slow machines (e.g. an older generation
smartphone) can notice that the performance of the game is negatively
impacted whenever advertisements are refreshed on the screen, to the
point the game cannot be played successfully. An expectation then
naturally arises, that these players should be able to play the game
on their platforms although this was not tested nor intended by Rovio
Entertainment.

The preferred process to overcome this inconvenience is either to
acquire a faster machine, or a license for the game that disables
advertisements lawfully, or convince Rovio to ensure compatibility
with the slower platforms. However, it is easy to conceive of players who are
unable and/or unwilling to go these routes before they experience the
gameplay.

Meanwhile, users have noticed that the game does not display
advertisements if it fails to connect to its advertisement server on
the Internet, in which case it can also deliver maximum
performance. Once understood, this effect is exploited by users by
purposefully disconnecting their device before starting the game.  The
system components (program \& device) are individually not modified;
this is why this change can be considered as extrinsic adaptation.

This example also reveals that a specific extrinsic adaptation
may be both desirable to a system's users and relatively undesirable
by the primary system providers. However, the example also reveals
that the existence of adaptation mechanisms may be a weighed trade-off
by the primary providers: while Rovio could in theory cause their
program to refuse to run if it cannot reach its advertisement server,
this would cause more user frustration and ultimately hurt their
bottom line more. Letting some users disable advertisements for the
sake of better game play is the lesser of two evils for the company.

\subsection{Fixing an incomplete specification in a video player}

We consider here the example of a video player consumer appliance
which can play movies from files stored digitally.  Typically, the end
users of such video players discover aspects of the running appliance
that were not considered initially by the makers. In this example, a
shortcoming in the specification of the menu navigation system may
render the appliance unable to play videos whose file name contain
non-ASCII characters. This observed behavior, and the corresponding
user expectation that this behavior should not occur (ie the player
should play the video nonetheless), is the symptom of a deviation
between the manufacturer's specification and the user-defined
specification. The situation is also described by saying that the
manufacturer's specification was incomplete.

The usual process for the user to reduce the deviation is to complain
to the manufacturer and request a firmware update or a refund to buy a
competing product without this defect. Both the user and the
manufacturer (and the state regulations) have agreed beforehand
explicitly that this process is available to both parties in case of a
problem.  However, this process yields a time to solution of a few
days to a few months, and this delay is usually not controllable by
the user. Any change that provides a stop-gap solution in a shorter
time frame is thus desirable to the end-user wishing to watch their
movie as soon as possible.

Meanwhile, a workaround also exists: for the user to rename the files to
using only ASCII characters. Furthermore, using the APIs
provided with most video player appliances nowadays, a
user-programmer is even able to implement an automated service that
performs this file name conversion automatically every time a new file
is added to the appliance's storage.

This workaround really tricks the existing components into matching
the user's requirement. It can be implemented without involving the
primary parties, namely the manufacturer and its design-supply
chain. This workaround is thus again an instance of extrinsic
adaptation.

This example highlights a key benefit and a key constraint for
extrinsic adaptation.  The key benefit is the ability for individual
parties to obtain a desired behavior more quickly and cheaply than
when relying only on the mechanisms and work flows envisioned by the
system providers. The key constraint is that extrinsic adaptation is
only possible if the system provides extension interfaces to alter its
behavior \textsl{a posteriori}, in this case the ability to rename
files.

\subsection{Compiler puppeteering for fast back-end prototyping}

This author has participated in a project that experimented with
different extensions to a processors's instruction set architecture
(ISA).

When changing the ISA of a processor, existing programming
language compilers cannot be reused as is. To program the new
processor, code generation from higher languages than assembly must be
implemented as well. However, traditionally code generators have been
implemented by considering ISA's as generally stable; the overall
process of creating a new ``back-end'' towards a new ISA is a
long-winded process in most compiler frameworks.

For this project it was thus necessary to find a solution that would
provide code generation in a faster way than the canonical process
offered by existing compiler frameworks. The technical solution
explained in~\cite[App.~H]{poss.12} is yet another extrinsic
adaptation, which encapsulates the GNU C compiler unchanged into a
wrapper program: the wrapper instruments each input source file, and
modifies its output assembly code. From the user's perspective, the
result looks like a new compiler supporting an
extended input language (SL~\cite{poss.12.sl}) and a different output
ISA. This wrapper is also considerably simpler to modify towards a new
target ISA; it was successfully ported to four different target ISAs
in less than a man-year worth of work, which would have been more
challenging to achieve by performing an intrinsic adaptation of an
existing compiler.

In this example, two key features of the compiler tool chain were
enabling factors for extrinsic adaptation. The first is that the
interface of a compiler, from the perspective of compiler users and
building tools, is standardized and ready to accept alternate
implementations. The second is that the GNU C compiler itself provides
extension mechanisms in its interface, namely run-time parameters to
code generation (\eg \texttt{-ffixed-reg}), the C preprocessor which
can be used to add new constructs in the language, and the
general-purpose \texttt{asm} statement which propagates any text from
a program to the ISA's assembler without checking its validity.

In other words, the extrinsic adaptation in this case is enabled by
both clean component interfaces and built-in mechanisms in the main
component to adapt its behavior via special inputs or parameters at
run-time.

\section{Extrinsic \vs invasive adaptation}\label{sec:extinv}

The definition in \cref{sec:def} insists on a view where a system is a composite
of components with well-defined boundaries, and that extrinsic
adaptations are limited to changing the composite without changing the
components themselves.

Component boundaries are crucial to evaluate the impact of a change
and, by contrast, how much of the rest of the system is left
unaffected. The concern here is to provide confidence that the
extrinsic adaptation does not remove support for previous requirements
(expected features). If the system is viewed as an
uncompartimentalized whole without clear component boundaries, or if
the boundaries are not respected and components are modified, no
confidence could be derived that a modification on a part (or adding a
part) has no impact on the behavior of the other parts.

Extrinsic adaptation is thus based on the definition of internal
component boundaries, so that a change to the system can be
objectively determined to occur either ``within'' or ``outside'' a
component.  If there are no internal boundaries, any change whatsoever
has potentially an impact on the entire system; we then call it
\emph{invasive}. In general, ``extrinsic'' \vs ``invasive'' are
extremes on a scale, and two changes can be compared according to
whether they are ``more extrinsic'' or ``less invasive'' than the other.

The reason why the distinction between invasive and extrinsic matters
is that extrinsic adaptations are subjectively considered more
\emph{acceptable} than invasive adaptations in the eye of (human)
external observers. \emph{In other words, when a secondary party
  changes a system and can argue that the change is more extrinsic, it
  increases its subjective value.} The value can be either monetary,
social, moral, ethical or philosophical, or any combination thereof.

A real-world example that highlights the distinction between extrinsic
and invasive adaptation can be found by considering a system
originally constituted by a horse and a person, and the addition of a
new requirement that the person should be able to ride the horse while
it is running, without falling. One approach to address the new
requirement is to perform a surgical operation to modify the bone
structure of both, so that the person's hip can firmly position
itself on the horse's back. Another solution is to add and use a
saddle.

There are multiple ways for an external observer to determine which
change is preferable at a subjective level.

One way is to argue that the surgical operation is invasive, because
it disrupts biological metabolism and may interfere with the
components' biological integrity in the longer term. In contrast, the
addition and use of a saddle respects the biological boundaries and
does not interfere with biological integrity. This argument makes the saddle
adaptation more acceptable by presenting it as an extrinsic
adaptation.

Another way is to argue that the saddle is invasive, because it breaks
the direct skin-to-skin contact between human and horse. This
separation can be argued to interfere with the overall reactivity of
the human-horse system to external stimuli. In comparison, the bone
structure operation preserves this physical contact. In this
perspective, the relationship between haptic feedback and reactivity
is central, and a saddle is more invasive thus less acceptable.

\section{Limited scope of cost functions}\label{sec:cost}

The previous section has highlighted that when component
boundaries in a system are unclear, any change to the system by a
secondary party may be either considered as extrinsic or invasive
depending on subjective criteria.

The distinction is often irrelevant in practice when changes can be
associated to a neutral \emph{cost function}. When available, a cost
function can both be used to determine whether a change is acceptable,
\ie when the cost fits within a \emph{budget}, and which of two
possible changes is preferable, \ie which has the \emph{lowest cost}.
The definition of clear-cut, objective cost functions may thus be seen
as a way to bypass the subjectiveness of extrinsic \vs invastive
adaptations altogether, and to provide an objective way to evaluate
system changes in general.

Alas, the subjectiveness cannot be eradicated entirely.

First, the audience may not agree on which cost functions to use and
their relative priority. A typical example is the evaluation of change
to a computing system with regard to implementation effort and
performance. If the audience does not agree on whether to use effort or
performance or both, the cost functions cannot serve to evaluate
changes. Then even if the audience agrees to consider both, with the
intent to use Pareto efficiency to evaluate changes, still no choice
can be made if the relative weight of each cost function is not
determined.

More fundamentally, cost functions are incomplete. When \emph{budget is
  not known \textsl{a priori}}, a cost function cannot determine
whether a change is acceptable overall. When two candidate changes
have \emph{equivalent cost}, cost alone cannot be used to pick one
over the other.

\begin{figure}
\hspace{-.2\textwidth}\includegraphics[width=1.4\textwidth]{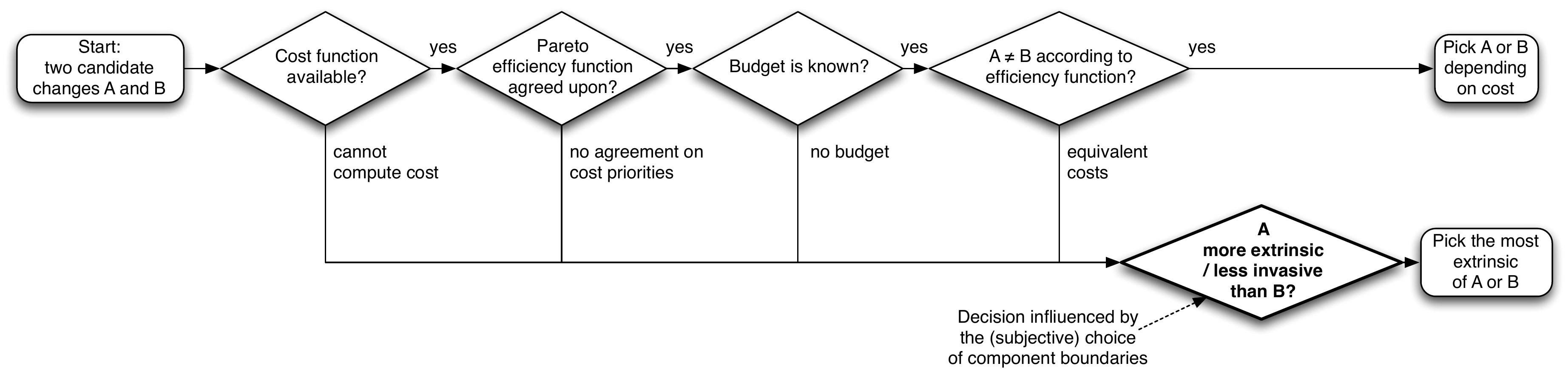}
\caption{Evaluation process of candidate adaptations to a system.}\label{fig:changecomp}
\end{figure}

We represent this situation in \cref{fig:changecomp}: whether a change
is invasive of extrinsic is the \emph{ultimate fallback arbitration}
when choosing change strategies. This fallback arbitration is also
subjectively influenced by the choice of component boundaries, as
highlighted previously. Cost functions cannot entirely
side-step this subjectiveness.

\section{Acceptability \vs choice of boundaries}\label{sec:acc}

We have highlighted in \cref{sec:extinv} that the determination of how
extrinsic/invasive a change is requires to first determine component
boundaries. If component boundaries are unclear or undefined,
evaluation is not possible.

However, any given choice of component boundaries may not be
sufficient either.  To understand this, consider that component
boundaries can be determined in different dimensions of human concern:
ownership, origin, vendor, physical frontier, licensing, etc.
Therefore, a given concrete change to a system can be more extrinsic from the
perspective of one boundary definition, and more invasive from the
perspective of another. This observation, together with
the observation above that extrinsic adaptations by secondary parties
are preferable to invasive adaptations, explains why the duality of
extrinsic \vs invasive adaptation creates irreconcilable differences
in human communities.

To illustrate this, we can consider the context of software systems,
where componentization commonly occurs in two spaces: engineering and
marketing. In the engineering space, behavior interfaces are used to
enable component-based software design. In the marketing space,
ownership and abstract feature boundaries\footnotemark{} are used to determine the
productization of the system.

\footnotetext{Database providers, for example, are
known to commonly introduce an artificial boundary between ``end users
seats'' and price licenses per seat, even when the technical
implementation only uses one database connection and server instance
for multiple end-users.}

In this context, there exists an ongoing struggle between proponents
of free software and organizations expecting to levy monetary income
from software integrations. The struggle occurs as follows. On the one
side, free software providers produce software components and systems
with the expectation that they can modify both the components and the
systems wherever they are reused by third parties.  On the other side,
organization reuse and integrate this software in larger systems
contained in a black box product boundary, either physical (\eg DRM) or legal
(\eg DMCA). From the free software provider's perspective, these new
boundaries are not relevant to componentization and they feel entitled
to adapt the system by adding new software components or modify the
integration of existing components. From their perspective, this adaptation is
extrinsic.  From the product provider's perspective, adding or
modifying the system necessarily violates the provider's physical or
legal boundaries, and is thus invasive. Depending on which side an
external observer stands on, the approach that changes such a system
by trespassing the physical/legal boundaries and changing the software
has either a high or low ethical/moral value.

\section{Measures of extrinsic adaptability}\label{sec:measure}

We assume that the \emph{ability} for secondary parties to perform
extrinsic adaptations to a system is generally \emph{desirable},
because it enables secondary parties to satisfy new requirements in a
shorter time or lower cost than relying on canonical updates by
primary parties.

This ability in turn depends on two factors: that extrinsic
adaptations are at all \emph{possible}, and that they can be perceived
as \emph{acceptable} according to objective (cost functions) and
subjective (invasiveness) criteria. Conversely, systems can oppose \emph{friction} against extrinsic
adaptations, either by:
\begin{itemize}
\item \emph{technical friction}: by making change difficult to implement;
\item \emph{friction against transparency}: by preventing the definition of
  component boundaries, so that external parties are pushed to
  consider changes to be generally invasive and thus undesirable.
\end{itemize}

Note that there is no friction specific to cost functions. Either cost
functions are readily available and agreed upon, in which case there
is no friction in this aspect; or primary parties make cost difficult
to identify or to agree upon, in which case evaluation of changes
falls back to the extrinsic/invasive scale subject to system
transparency.

In general, a general measure of \emph{extrinsic adaptibility} can be
taken by evaluating \emph{how little friction} a system opposes to
extrinsic adaptations.  The following sub-sections detail which
measures of friction can be used for this purpose.

\subsection{Technical friction}

There are three general ways that a system's design can oppose technical
friction to extrinsic adaptations:

\begin{itemize}
\item \emph{friction against alternate integration}: makes it difficult to
  combine the existing, unchanged components of the system in a
  different way, to disable/remove existing components altogether, or to
  provide them with different input parameters at run-time that could
  change their behavior;
\item \emph{friction against extension}: makes it difficult to add new
  components next to the existing components in the system;
\item \emph{friction against change resilience}: makes it difficult to keep
  using the adaptation over time.
\end{itemize}

We do not separate ``friction to substitution'', because substitution
is really the combination of adding a new component (extension), then
disable the existing component and adapt the system to use the new
component (alternate integration). Friction to either of these
sub-actions is sufficient to oppose substitution.

Note that there exist another form of technical friction against
change in general, although this form is not relevant to extrinsic
adaptations: friction to component modification. this occurs
when a design makes it difficult to open a black box component and
change part of its implementation while keeping the rest. This is not
a concern for extrinsic adaptation which, by definition, does not
involve component modifications and instead relies on changes to the
integration between components, assumed to be visible.

The mechanisms by which primary providers create technical friction to
extrinsic adaptations in a system include:
\begin{itemize}
\item \emph{tight integration}: using complex and
  multi-layered inter-dependencies between components, which generally
  makes it difficult to change one part of the system without altering
  the rest;
\item \emph{warranty seals}: refusal to deliver further service or updates
  after non-canonical changes;
\item \emph{component signatures}: cryptographic signing of component
  implementations, and constructed social disapproval of behaviors from
  components signed by untrusted sources.
\end{itemize}

Tight integration is a form of friction against alternate
integration. Warranty seals oppose all three forms of friction
identified above. Component signatures oppose friction against
extension, and thus also against substitution.

\subsection{Friction against transparency}\label{sec:transf}

With limited transparency, any secondary party wishing to develop an
adaptation first needs to investigate the system's inner workings and
reconstruct \textsl{a posteriori} a model of the interaction between
its components, \ie perform first an act of \emph{reverse engineering}
the system back to a design specification.

The mechanisms by which primary providers create friction against
transparency and reverse-engineering include:
\begin{itemize}
\item \emph{secrecy}: existing knowledge and documentation about the system's
  design is hidden away from secondary parties;
\item \emph{opaque physical barriers}: using a sealed physical
  enclosure (bolted enclosure,
  chip package, etc.) which prevents physical observation;
\item \emph{obfuscation}: addition of extra components
  and interactions which are not necessary to proper system function
  and just serve to increase the effort necessary for
  reverse-engineering;
\item \emph{encryption}: encryption of interactions
  between components, which hides the interaction protocols and
  prevents reverse-engineering component interfaces;
\item \emph{alienation}: using an unusual overall design for the system
  when alternate, better-known approaches exist.
\end{itemize}

\subsection{Relative impact of friction types}\label{sec:ftypes}

Friction against transparency is fundamentally less strong than
technical friction. Indeed, as soon as one sufficiently motivated
secondary party overcomes obstacles to transparency, the accrued
knowledge from reverse-engineering can be gathered, published and made
accessible for multiple subsequent adaptations. The costs incurred by
friction against transparency can thus be amortized over time.

In contrast, technical friction induces extra implementation costs in
\emph{each} extrinsic adaptation, and thus cannot be amortized. When found in
a system, friction against transparency can be perceived more as a mere temporary
inconvenience, whereas technical friction can be perceived as a
serious obstacle worth complaining about.

\section{Extrinsic adaptability in computing systems}\label{sec:compsys}

Software systems are traditionally well componentized. With the
popularization of open source software in the late 1990's, and the
large democratization of software engineering that followed, the friction
against extrinsic adaptation of software has been generally
reduced. Since then, despite the occasional transparency obstacle in
proprietary, closed source components, large software systems are
largely extrinsically adaptable.

The picture is not so clear when the boundary of a computing system is
extended to include hardware components, \ie the hardware
platform.

On the one hand, the industrial world where computing hardware is
produced is culturally largely opposed to transparency, and the
various mechanisms identified in \cref{sec:transf} are commonly
used. And in the occasional case where secrecy, obfuscation,
encryption and alienation are not expressely sought by system
providers, miniaturization and especially chip packages create opaque
physical barriers that are difficult to overcome. Even though friction
to transparency can be considered mostly an inconvenience (\cf \cref{sec:ftypes}), the associated overheads to extrinsic adaptation
exist in every hardware platform.

On the other hand, there exist large opportunities to reduce technical
friction, although they are currently not commonly
considered.

To begin with, consider that large efforts were made in the last three
decades of the 20th century to build modular hardware platforms, \eg with
interchangeable memory, processor and interconnect components. This
systematic reduction of technical friction against extrinsic
adaptation was justified by a strong demand for adaptable hardware by
the user base, which itself was a side-effect of the personal
computing revolution of the '70s and early '80s.

Since then, as the customer base for hardware products has extended
beyond the community of hobbyists, the demand for adaptable hardware
has \emph{relatively} dwindled\footnotemark{}. Simultaneously, tight sales margins
favor tight integration, often in the form of specialized hardware and
systems-on-chip. And for all systems with multimedia output, the
political atmosphere has facilitated the pervasive use of component
signatures. All these factors have created tremendous technical
friction against extrinsic adaptation.

\footnotetext{We postulate that in absolute economic value, the demand
  for adaptable systems has even increased, albeit less fast than the
  overall demand for computing systems.}

But the current situation does not invalidate the benefits and
desirability of extrinsically adaptable systems. A known way to design
for extrinsic adaptation is to exercise hardware modularity and
standard hardware interfaces. More adaptable systems would
reduce vendor lock-in, increase reusability and thus eventually reduce
waste and evolution costs. Moreover, renewed interest and demand for
extrinsic adaptability is only a slight cultural shift away: a more
open political mindset, and a manufacturing overhead eventually
compensated by the aforementioned longer-term benefits.

We can thus argue for the existence of an \emph{ethical imperative}
for hardware architects and platform providers to stimulate extrinsic
adaptivity, as a particular instance of the more general imperative to
favor long-term priorities over short-term gains.

\section{Conclusion}\label{sec:conc}

We have distilled the concept of \emph{extrinsic adaptation} to a system
of components, a generalization which encompasses the popular concepts
of ``workaround'' or ``hack'' when talking about changes to an
existing system by secondary parties.

We have also highlighted that system changes by secondary parties fall
on a subjective gradient of acceptability, with extrinsic adaptations
on one side which confidently preserve existing system features, and
invasive modifications on the other side which are perceived to be
disruptive to existing
system features. Where a change is positioned on this gradient is
dependent on how an external observer perceives component boundaries
within the changed system. We have also argued that the existence of
objective cost functions can alleviate but not fully eliminate this
subjectiveness.

The general benefits of extrinsic adaptations have then led us to
define general metrics to evaluate how much systems are amenable to
extrinsic adaptation, \ie measure their extrinsic adaptability. We do
this by considering the opposite force, \ie friction against
adaptability, and identifying the mechanisms currently in use by
system providers to oppose friction.

We have then applied the proposed concepts to the case of computing
systems, and formulated a general ethical imperative to promote
extrinsic adaptability in hardware platforms.

\section*{Acknowledgements}
\addcontentsline{toc}{section}{Acknowledgements}

The author is grateful to Sebastian Altmeyer, Michiel W. van Tol and
Roy Bakker for their helpful comments and suggestions.

\nocite{poss.13.coord}

\newcommand{\etalchar}[1]{#1} % pre-defined so the bst does not complain
\addcontentsline{toc}{section}{References}
\bibliographystyle{is-plainurl}
\bibliography{doc}

\end{document}